\documentclass[prl,aps,amsmath,twocolumn,floatfix,showpacs]{revtex4}
\usepackage{epsfig}

\begin{document}

\title{Quantized phase shifts and a dispersive universal quantum gate}
\author{F.~Schmidt-Kaler, H.~H\"affner, S.~Gulde, M.~Riebe, G.~Lancaster, J.~Eschner, C.~Becher,
 and R.~Blatt}

\affiliation{Institut f{\"ur} Experimentalphysik, 6020 Innsbruck,
Austria}

\date{\today}

\begin{abstract}
A single $^{40}$Ca$^+$-ion is trapped and laser cooled to its motional ground
state. Laser radiation which couples off-resonantly to a motional sideband of
the ion's $S_{1/2}$ to $D_{5/2}$ transition causes a phase shift proportional
to the ion's motional quantum state $|n\rangle$. As the phase shift is
conditional upon the ion's motion, we are able to demonstrate a universal
2-qubit quantum gate operation where the electronic target state $\{S,D\}$ is
flipped depending on the motional qubit state $|n\rangle=\{|0\rangle,
|1\rangle\}$. Finally, we discuss scaling properties of this universal
quantum gate for linear ion crystals and present numerical simulations for
the generation of a maximally entangled state of five ions.
\end{abstract}

\pacs{PACS number(s): 03.67.Lx,03.65.Ud}

%03.67.Lx Quantum computation
%03.65.Ud Entanglement and quantum nonlocality (e.g. EPR paradox, Bell's
%inequalities, GHZ states, etc.)
%32.80.Qk   Coherent control of atomic interactions with photons

 \maketitle

%Einleitung

%Spektroskopie und Pulsschema beschreiben
%Bild 1: a) Leiter schema, b) Messmethode als Puls Schema
%Bild 2: Der gesamte Datensatz quantisierten Stark-Shifts

%Erklärung des Gatters
%Bild 3: 4 Rabiosz. als Wahrheitstabelle des Gatters

%Zukunfts-Musik für 5 Ionen
%Bild 4: num Sim. SSSSS+DDDDD

%Finale

Trapped ions are successfully used for quantum information processing
\cite{PhysQI00,CHUANG00,ARDA} as experiments with a single ion
\cite{MONROE95,ROOS99,GUL03} and two-ion crystals \cite{LEIB03,SCH03} have
proven. Even four-ion crystals \cite{SACK00} have been transferred into an
entangled state. For the implementation of quantum algorithms, single- and
two-qubit operations are required \cite{CHUANG00}. For most implementations
of two-qubit gate operations, a well defined phase shift is induced,
depending on the control qubit's quantum state. Its origin might be either a
geometric phase \cite{SJOK00,DUAN01} as the quantum state follows a closed
loop in phase space \cite{LEIB03}, or the sign of the wave function is
changed as a resonant $2\pi$-Rabi rotation is performed
\cite{CIR95,MONROE95,SCH03}. Alternatively, the phase shift which is required
for the gate operation may be derived from a non-resonant laser-atom
interaction. The resulting quantum phase gate $\Phi$ transforms the state of
two qubits $|a,b\rangle \rightarrow \exp{(i \phi \delta_{1,a} \delta_{1,b})}
|a,b\rangle$, where $\delta_{1,a}$ and $\delta_{1,b}$ denote Kronecker
symbols with $\delta_{1,a}=1$ if qubit $a$ is in the state $|1\rangle$ and
$\delta_{1,a}=0$ otherwise. Only for both qubits in $|1\rangle$, the quantum
state's phase is shifted by $\phi$. This phase gate may be converted into a
controlled-NOT operation (CNOT) by single qubit rotations before and after
performing $\Phi$ to yield $|a,b\rangle \rightarrow |a,a \oplus b\rangle$,
where $\oplus$ represents an addition modulo 2.

A quantum phase gate \cite{RAUSCH99} also has been demonstrated using long
lived superpositions of Rydberg atoms which traverse a high-Q microwave
cavity. The phase $\phi$ of a superposition $(|0\rangle
+\text{e}^{i\phi}|1\rangle)/\sqrt{2}$, where $|0\rangle$ and $|1\rangle$
denote the qubit states, is changed conditioned on the quantum state of the
cavity mode. If the cavity resonance is detuned slightly from that of the
atomic qubit transition, superpositions acquire a significant light shift
already by a single photon stored in the microwave cavity \cite{BRUNE94}.

%Another approach utilizes nonlinear phase shifts
%arising from the birefringence of a single atom strongly coupled to a
%high-finesse optical resonator \cite{TURCH95}.
%
%%Here, single photons carry the qubit information and their
%%polarization state is changed conditioned on the intracavity
%%photon number.
%
%\textbf{????? Similar attempts have been performed in cavity QED
%experiments with optical fields \cite{TURCH95}.}

% the measurement of the Wigner function of the photon Fock state
%$|n=1\rangle$ \cite{NOGUES00} and the quantum-non-demolition measurement of a
%single photon \cite{NOGUES99} has been realized. Similar attempts have been
%performed in cavity QED experiments with optical fields \cite{TURCH95}.

\begin{figure}[t,b,h]
\begin{center}
\epsfig{file=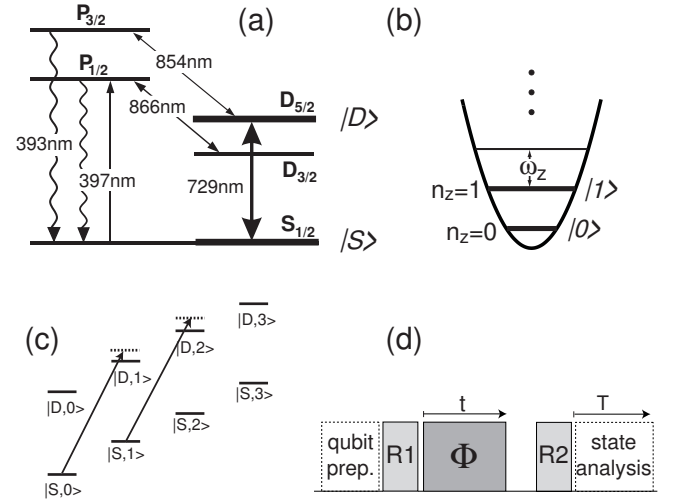,width=.99\linewidth} \caption{\label{Levels&Sequence}
a) $^{40}$Ca$^+$ level scheme. A qubit is encoded in the $S_{1/2}, (m=-1/2)$
ground and $D_{5/2}, (m=-1/2)$ metastable state. b) The lowest two number
states $n$ of the axial vibrational motion in the trap form the other qubit.
c) Non-resonant interaction with the laser tuned close to the upper motional
sideband (indicated by arrows). d) Laser pulse sequence for the quantum gate,
consisting of the preparation of qubit states, single bit operations R1 and
R2 and controlled phase shift operation $\Phi$ of duration $t$. Finally the
qubits are analyzed for their electronic and motional quantum state.}
\end{center}
\end{figure}

In this paper we study experimentally and theoretically the dispersive
(off-resonant) coupling of laser radiation for quantum information processing
with trapped ions. The quantum information is encoded in the ion's electronic
ground state $S_{1/2}, (m=-1/2) \equiv |S\rangle$ and a long lived metastable
state $D_{5/2}, (m=-1/2) \equiv |D\rangle$ (see Fig.~\ref{Levels&Sequence}a).
The qubit can be manipulated by laser radiation near 729~nm on the
corresponding quadrupole transition.
% In this first part of the paper we describe how to induce and to
%determine a phase shift of
%a single qubit in the superposition $|S+D\rangle/\sqrt{2}$  depending on the
%Fock state in which the motional state is prepared (see
%Fig.~\ref{Levels&Sequence}a, indicated by natural numbers).
The motional state of the ion is prepared in a Fock state (see
Fig.~\ref{Levels&Sequence}b) and serves as control qubit. The interaction
with off-resonant laser light (see Fig.~\ref{Levels&Sequence}c) induces a
phase shift of the target qubit (electronic states) conditioned upon the
ion's motional state.
% The measurement
%principle is illustrated in Fig.~\ref{Levels&Sequence}b.
In our experiment, a single $^{40}$Ca$^+$ ion is stored in a linear Paul trap
with a radial and axial frequency of $\omega_{rad}/2\pi=4.9$~MHz and
$\omega_{ax}/2\pi=1.712$~MHz, respectively. Doppler cooling, followed by
sideband ground state cooling and optical pumping leads to a population of
the state $S_{1/2}, (m=-1/2) \equiv |S\rangle$ and its lowest motional
quantum state $|n=0\rangle$ with more than 98$\%$ probability. The internal
state is detected by electron shelving \cite{ROOS99} with an efficiency
$>99\%$.
%For this we irradiate the ion with light near 397~nm on the dipole
%allowed $S_{1/2}$ to $P_{1/2}$ transition and project thus its quantum state
%onto the computational basis $\{S,D\}$; if the ion is found in $|S\rangle$,
%we detect its fluorescence light on a photomultiplier. For the $|D\rangle$
%state, no light is observed.
We determine the occupation probablity of the $|D\rangle$ state, $P_D$, by
averaging over 100 experimental sequences.
% Averaged over 100 experimental
%sequences, we reveal the probability $P_D$.
 Details of the cooling method and
the experimental apparatus are described elsewhere~\cite{HAEFF03,ROOS99}.
Starting from the state $|S,n=0\rangle$, laser pulses resonant to the blue
axial sideband $(|S,n\rangle \leftrightarrow |D,n+1\rangle)$ and resonant to
the carrier transition $(|S,n\rangle \leftrightarrow |D,n\rangle)$ of the
qubit transition are used to transfer the motional state to Fock states
\cite{ROOS99}. The transfer quality of blue sideband $\pi$-pulse and carrier
$\pi$-pulse exceed $98\%$ and $99\%$, respectively. Thus we yield the desired
motional Fock state with fidelities better than $96\%$, $94\% $ and $92\% $
for $n$=1, 2 and 3, respectively.

The phase evolution of the atomic wavefunction due to interaction with
non-resonant laser radiation is probed in a Ramsey experiment: A resonant
$\pi/2$ pulse ($R1$ in Fig.~\ref{Levels&Sequence}d) on the quadrupole
transition prepares the the electronic qubit in the superposition state
$(|S\rangle +|D\rangle)/ \sqrt{2}$. A second resonant $\pi/2$ laser pulse,
$R2$ is applied with opposite laser phase and a delay of $t=260~\mu$s. If the
atomic wavefunction does not acquire any additional phase during the delay
time, $R2$ would undo the operation of $R1$ and transfer the electronic state
back to $|S\rangle$.
\begin{figure}[t,b]
\begin{center}
\epsfig{file=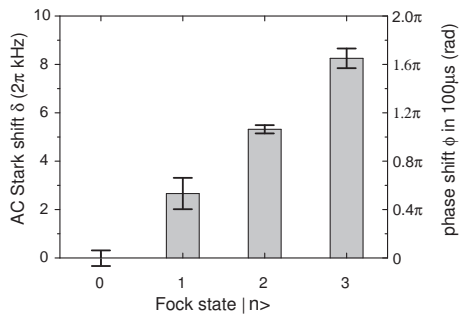,width=.7\linewidth} \caption{\label{123_phonon}
Residual AC Stark shift (left axis) and the corresponding phase shift of the
wavefunction (right axis) for states with up to $n=3$ axial phonons. A linear
fit yields $\delta_{\text{res}}/2\pi=2.71(6)$~kHz per phonon. Here, the phase
shifting laser is detuned from the blue axial sideband by $\Delta=60$~kHz.}
\end{center}
\end{figure}
For generation of a deterministic phase shift during the delay time, the ion
is exposed to laser light which is blue-detuned by $\Delta$ from the blue
axial sideband of the qubit transition. The resulting phase shift is due to
i) off-resonant coupling to dipole transitions (see
Fig.~\ref{Levels&Sequence}a), ii) off-resonant coupling to carrier
transitions of the $S_{1/2}$ to $D_{5/2}$ Zeeman manifold and iii)
off-resonant coupling to the motional sidebands of the qubit transition. An
extensive study of the first two contributions is given in \cite{HAEFF03}.

The relative strengths of the three contributions scale with their respective
Rabi frequencies. Thus, the light (AC Stark) shift $\delta_{(\text{iii})} =
\Omega_{n,n+1}^2/4\Delta$ due to the third contribution is weaker by almost
two orders of magnitude compared with (i) and (ii), as the resonant Rabi
frequency of the blue sideband between $|S,n\rangle$ and $|D,n+1\rangle$ is
given by
\begin{equation}
  \Omega_{n,n+1}=\eta \hspace{0.5mm} \Omega_0 \hspace{1mm}
\sqrt{n+1} \label{rabiBS}
\end{equation}
with the Lamb-Dicke factor $\eta=0.068 \ll 1$. As $\delta_{(\text{iii})}$
strongly depends on the motional quantum number it can be utilized to
generate conditional phase shifts. However, the stronger contributions (i)
and (ii) mask this effect. Therfore we cancel (i) and (ii) using an
additional off-resonant light field with equal, but opposite phase shift
("compensating" light field) which illuminates the ions during the phase
shift operation \cite{HAEFF03}. This light field is derived from the same
laser source near 729~nm which we use for the qubit transition using an
acousto-optical modulator. By optimizing the compensation light field
parameters, i.e. laser intensity and detuning,
we also cancel the light shift %$\Delta \phi = \delta_{(iii)} t$ during the
%delay time $t$
for the motional Fock state $|n=0\rangle$. For this state the compensation
leads to a residual light shift of $\delta_{\text{res}}/2\pi \leq 300$~Hz
(see Fig.~\ref{123_phonon}). Thus, after compensation, the phase shift
$\Delta \phi = \delta_{\text{res}} t$ is directly proportional to the
motional quantum number $n$ of the ion,
\begin{equation}\label{phaseShift}
  \Delta \phi = \frac{\eta^2 \Omega_0^2 \hspace{1mm} t}
  {4 \Delta}\hspace{1mm} \hspace{1mm} n
\end{equation}
where $t$ denotes the duration of the phase shift operation in the sequence.
For the Fock states $|n=1\rangle$ to $|n=3\rangle$, we observe a linear slope
of $\delta_{\text{res}}/2\pi=2.71(6)$~kHz per phonon (see
Fig.~\ref{123_phonon}).

In order to realize a phase gate operation, we chose $t=t_0$ such that the
residual phase shift for states with $n=1$ yields precisely $\pi/2$. This
accomplishes a quantum gate in the two-qubit computational space composed by
the states $|S,0\rangle$, $|D,0\rangle$, $|S,1\rangle$ and $|D,1\rangle$: The
states $|S,1\rangle$ and $|D,1\rangle$ acquire phases of $e^{\pm i\pi/2}=\pm
i$, respectively, while $|S,0\rangle$ and $|D,0\rangle$ do not acquire any
phase. Thus the phase shift operation $\Phi$ reads:
\[
\Phi(t_0)=\left(\begin{array}{cccc} 1&0&0&0\\
0&1&0&0\\0&0&i&0\\0&0&0&-i\end{array}\right).
\]
%Moreover, as the laser field of pulse $R2$ is adjusted to have a
%phase of $\pi$ with respect to $R1$, the corresponding matrices
%read where the $+$ signs refer to R1 and the $-$ to R2.
The Ramsey pulses $R1$ and $R2$ are represented by
\[
\frac{1}{\sqrt{2}}\left(\begin{array}{cccc} 1&\pm i&0&0\\
\pm i&1&0&0\\0&0&1&\pm i\\0&0&\pm i&1\end{array}\right),
\]
where the + (-) sign refers to $R1$($R2$). The full sequence $C$ represents a
universal two-qubit gate:
\[
C= \hspace{1mm} R2 \hspace{2mm} \Phi(t_0) \hspace{2mm} R1=\left(\begin{array}{cccc} 1&0&0&0\\
0&1&0&0\\0&0&0&-1\\0&0&1&0\end{array}\right).
\]
\begin{figure}[t,b]
\begin{center}
\epsfig{file=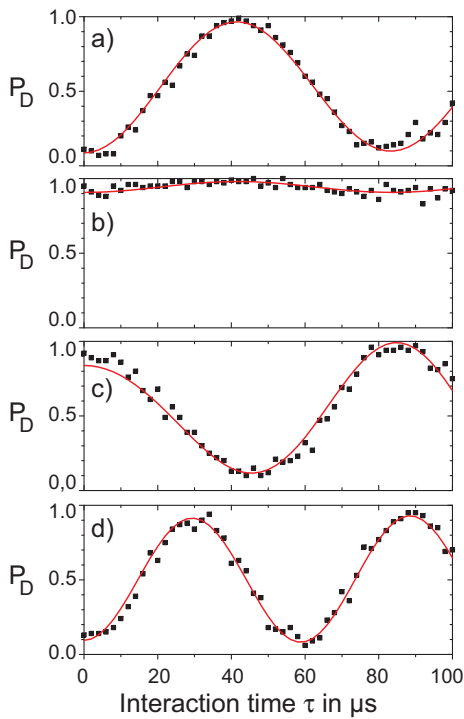,width=.7\linewidth} \caption{\label{TruthTable} State
analysis after the gate operation for input states $|S,0\rangle$ (a),
$|D,0\rangle$ (b), $|S,1\rangle$ (c) and $|D,1\rangle$ (d). The ion interacts
on the blue sideband ($|S,n\rangle \leftrightarrow |D,n+1\rangle$) resonant
for a time $\tau$.  A fit of pure sine-functions yields Rabi frequencies of
$\Omega_{0,1}/2\pi$=11.9(1)~kHz (a), 12.0(2)~kHz (c) and
$\Omega_{1,2}/2\pi$=16.95(5)~kHz (d). The frequency ratios of (d) and (a):
$\sqrt{2.02(4)}$, and of (d) and (c): $\sqrt{1.99(7)}$ agree well with the
expected factor $ \sqrt{2} $ from eq.~\ref{rabiBS}. The solid lines
correspond to a fit of the model function eq.~\ref{fit}.}
\end{center}
\end{figure}
In the experiment, the truth-table of this gate is probed by preparing the
four computational basis input states. %(using appropriate combinations of
%resonant carrier and resonant sideband $\pi$-pulses in the first period of
%sequence Fig.~\ref{Levels&Sequence}b).
After the gate %$(R1 \hspace{1mm} \Phi\hspace{1mm} R2)$,
operation $C$ the internal state $\{S,D\}$ which served as the target qubit
is detected by electron shelving. For detection of the motional state
$n=\{0,1\}$, i.e. the control qubit, we drive Rabi oscillations on the blue
axial sideband. % and infer $n$ from eq.~\ref{rabiBS}. %and infer from its frequencies the motional state $|n\rangle$
%according to equ.~\ref{rabiBS}.
These Rabi oscillations are plotted in Fig.~\ref{TruthTable}. In a
first step we determine the Rabi frequencies by fitting the data
with pure sine-functions. From this fit we determine
$\Omega_{0,1}/2\pi$=11.9~kHz and $\Omega_{1,2}/2\pi=\sqrt{2}
\Omega_{0,1}$ (c.f. Fig.~\ref{TruthTable}). In a next step we fit
the respective experimental data in Fig.~\ref{TruthTable}~a-d with
a model function:
\begin{eqnarray} P_D(\tau)&=&a_{S0} \sin^2(\Omega_{0,1}
\tau) + a_{D0} + a_{S1} \sin^2(\Omega_{1,2} \tau) +\nonumber
\\ & &  a_{D1} \cos^2(\Omega_{0,1} \tau) \label{fit}
\end{eqnarray}
The coefficients $a_{S0}, a_{D0}, a_{S1}, a_{D0}$ account for the
contributions of the four computational basis states $|S,0\rangle,
|D,0\rangle, |S,1\rangle, |D,1\rangle$ and obey
$a_{S0}+a_{D0}+a_{S1}+a_{D1}=1$. Table~\ref{TruthTableTable} lists these
coefficients for the four different input states. We find transfer
efficiencies between the ground state $|S,0\rangle$ and the desired output
states of $\geq81\%$. The measured efficiencies are attributed to the
following experimentally determined imperfections: (i) state preparation as
mentioned above (ground state cooling and preparation pulses) (ii) the chosen
gate interaction time $t_0=200~\mu$s which has been off by $+~8\%$ from its
ideal value, (iii) the independently measured loss of coherence ($\simeq
6\%$) of the qubit \cite{SCH03b} within the delay time between $R1$ and $R2$
and (iv) off-resonant sideband excitations during $\Phi$ \cite{STEA00},
measured here to account for an error of 4\%.
\begin{table}
\caption{\label{TruthTableTable} \rm{Experimentally obtained truth table of
the gate including the imperfect input state preparation.}}
 \center
\begin{tabular}{|c||c|c|c|c|}
\hline $\downarrow$input&$|S,0\rangle$&$|D,0\rangle$&$|S,1\rangle$&$|D,1\rangle$\\
\hline\hline $|S,0\rangle$&\textbf{0.90(1)}&0.06(1)&0.01(2)&0.03(1)\\
\hline $|D,0\rangle$&0.09(1)&\textbf{0.89(1)}&0.00(1)&0.02(1)\\
\hline $|S,1\rangle$&0.00(1)&0.03(1)&0.16(2)&\textbf{0.81(2)}\\
\hline $|D,1\rangle$&0.07(1)&0.00(1)&\textbf{0.84(2)}&0.09(2)\\
\hline
\end{tabular}
\end{table}

The demonstrated dispersive two-qubit gate can be extended to a larger number
of $N$ qubits. An important advantage of dispersive gates, like e.g. the gate
presented in ref.~\cite{LEIB03}, is that the computational subspace is
conserved automatically as the off-resonant interaction $\Phi$ only modifies
the phase. Other gate schemes need auxiliary levels \cite{MONROE95,CIR95} or
composite pulse techniques \cite{GUL03} thus increasing the technical
complexity of the experimental sequence.
%For certain
%computational protocols, it may be of advantage to use the dispersive gate:

\begin{figure}[t,b]
\begin{center}
\epsfig{file=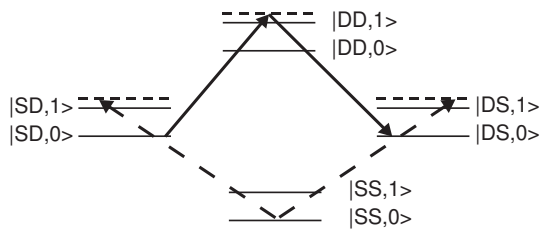,width=.9\linewidth} \caption{\label{twoionlevels} Level
scheme of a two ion crystal up to phonon number of $n=1$. The virtual levels
detuned from the blue sideband are indicated as dashed horizontal lines. The
solid arrows represent the Raman-like coupling between $|SD´,0\rangle$ and
$|DS,0\rangle$. The dashed arrows symbolize the off-resonant phase changing
coupling of the $|SS,0>$ levels on the blue sideband. }
\end{center}
\end{figure}

For the entangling operation, we select a symmetric vibrational mode (bus
mode) where the absolute values of all the ions' Lamb-Dicke parameters are
equal, e.g. the center of mass mode. Similar to the one-ion gate operation,
the $N$-ion string is assumed to be cooled close to the motional ground state
(e.g. by EIT-cooling \cite{FSK01}). To create an entangled state, we use a
pulse sequence $R1 \Phi(t_0) R2'$ acting in the following way: $R1 \Phi(t_0)
R2' |SS\ldots S \rangle = |SS\ldots S \rangle + \text{e}^{i\phi} |DD\ldots D
\rangle$. First we consider the two ion case. The two ions are excited by a
$R1$ pulse as defined above yielding the superposition state $1/2\left(|SS
\rangle+|SD \rangle+|DS \rangle+|DD \rangle\right)$. The phase operation
$\Phi(t_0)$ then couples the ions via the bus mode. The phase of the final
$\pi/2$ Ramsey pulse $R2'$ has to be chosen $\pi (\pi/2)$ relative to $R1$
for an odd (even) number of ions. Note that this scheme only requires the
same Rabi frequency for all ions, but no individual addressing of the ions.
This can be conveniently achieved by illuminating the ions along the trap
axis. The crucial part of the operation is contained in $\Phi(t_0)$: The
fraction of the superposition being in the $|DD \rangle$-state is dark and
thus remains completely unchanged. Similary to the one-ion case, the
$|SS\rangle$-part, however, acquires a phase factor without a change in
population. Choosing the pulse length $t_0$, Rabi frequency $\Omega$ and
detuning $\Delta$ of the sideband interaction such that $t_0=\Delta / (\eta^2
\Omega^2)$ the acquired phase factor of the $|SS\rangle$-state is $-1$. For
ions where both qubit levels are populated, Raman processes take place (see
Fig.~\ref{twoionlevels}) where the populations of the $|SD\rangle$ and
$|DS\rangle$ are exchanged and a phase factor of $-1$ is acquired.
The entangling mechanism is reminiscent of the
M{\o}lmer-S{\o}rensen scheme which, however, uses a non-resonant
bichromatic light field \cite{MOE99}.

A similar reasoning holds for $N$ ions as verified by numerical simulations.
Taking $N=5$ as example, we chose the parameters $\Omega=2\pi\times 230$~kHz,
$\Delta=2\pi\times60$~kHz, and $\eta=0.068/ \sqrt{5}$. In addition to the bus
mode (center of mass) the two closest spectator modes were taken into
account. After an interaction time of 900~$\mu$s a state
\begin{eqnarray}
\label{sssss+ddddd} |\Psi\rangle & = & \sqrt{0.48}
|SSSSS,000\rangle+\text{e}^{\i\phi} \sqrt{0.45}|DDDDD,000\rangle \nonumber \\
& &+\sqrt{0.07}|\epsilon\rangle
\end{eqnarray}
 is achieved, were
$|\ldots,n_b n_{s_1} n_{s_2}\rangle$ refers to the phonon numbers of the
bus-mode and two of the spectator modes, respectively. $|\epsilon\rangle$ is
a superposition of all undesired states with %a total population of
%$|\epsilon\rangle$ is 0.07 and has also
mostily zero phonon excitation. The population in $|\epsilon\rangle$ could be
reduced further by reducing the gate speed. The simulations take into account
most of our current experimental parameters and imperfections.

A spin-echo technique reduces the influence of frequency fluctuations which
are slow compared to the entangling time. Applying this technique, we observe
an interference contrast of more than 0.9 over 2~ms with a single ion. Thus,
we find that --for our current experimental imperfections-- it is reasonable
to entangle 5 ions with fidelities of about 80$\%$. Unlike most other
proposals the entangling time is $\sim \sqrt{N}$ and works for any number of
ions.

In conclusion, we have demonstrated a dispersive quantum gate operation
employing light shifts conditional on the vibrational quantum number in a
single ion. With numerical simulations, we explore the scheme for larger ion
numbers and propose the generation of maximally entangled states for ion
crystals (Eq.~\ref{sssss+ddddd}) with high fidelities.

\end{document}